\documentclass[amsmath,superscriptaddress,prl,twocolumn] {revtex4}
\usepackage{bm}
\usepackage{enumerate}
\usepackage{xcolor}

\usepackage{graphicx}
\usepackage[dvips]{epsfig}
\usepackage{epsf}

\makeatletter
\newcommand{\Rmnum}[1]{\expandafter\@slowromancap\romannumeral #1@}
\makeatletter

\begin{document}

\title{Magnon equilibrium spin current in collinear antiferromagnets}

\author{Vladimir A. Zyuzin}
\affiliation{L.D. Landau Institute for Theoretical Physics, 142432, Chernogolovka, Russia}

\begin{abstract}
We theoretically predict that the Dzyaloshinskii–Moriya interaction can induce a magnon equilibrium spin current in collinear antiferromagnets. Such a current, being a response to the effective magnon vector potential, can be considered the magnon analog of the supercurrent in superconductors. The large amplitude of the predicted effect may compensate for the smallness of the Dzyaloshinskii–Moriya interaction, making the equilibrium spin currents experimentally observable.
Importantly, we show that the current being the ground state property isn't zero at zero temperature.
We suggest that an external electric field can play the role of an effective flux that magnons interact with, and we propose an experiment based on the interference of magnons in a ring geometry to verify the concept.
\end{abstract}
\maketitle

Transport properties of magnons in collinear antiferromagnets have been the subject of research in the past decade \cite{AFMspintronicsRMP,MatsumotoShindouMurakamiPRB2014,ZyuzinKovalevPRL2016,Naka2019,HayamiYanagiKusunose2019,Exp2026,Zyuzin2025b}. 
Magnons are neutral quasiparticles, and they can transport either spin or heat in response to an applied temperature gradient. Microscopic theory of the thermal Hall and magnon spin Nernst effects in collinear antiferromagnets as a response to a temperature gradient has been developed in \cite{MatsumotoShindouMurakamiPRB2014,ZyuzinKovalevPRL2016,Zyuzin2025b}. These are dissipationless responses and are defined by the Berry curvature of magnons in the Brillouin zone.
Dissipative anisotropic magnon spin currents (spin-splitter effect) in collinear antiferromagnets due to spin splitting have been recently theoretically predicted \cite{Naka2019,HayamiYanagiKusunose2019} and experimentally confirmed \cite{Exp2026, AlterComment}. Such spin splitting occurs in subclasses of compensated N\'{e}el orders belonging to the symmetry classes of PT or translation and time-reversal symmetry broken antiferromagnets, such as weak ferromagnets, piezomagnets, or magnetoelectrics \cite{Naka2019,HayamiYanagiKusunose2019,Zyuzin2025b,BorovikRomanov1957,Dzyaloshinskii1958, Moriya1960b, Turov1965,BorovikRomanov1960,BorovikRomanov1973, VonsovskiiTurov1986, Turov1990,Krichevtsov1981,KharchenkoBibikEremenko1985,EremenkoKharchenko1987, TurovShavrov,Solovyev1997} and ferrimagnets \cite{Zyuzin2025b}.
Dissipative anisotropic magnon spin currents \cite{Naka2019,HayamiYanagiKusunose2019} are different from the magnon spin Nernst effect \cite{ZyuzinKovalevPRL2016}. If the magnon spin Nernst effect is the analog of the Hall effect in fermion systems, then the dissipative anisotropic magnon spin currents obtained in \cite{Naka2019,HayamiYanagiKusunose2019} are the analog of the linear d-wave magnetoconductivity discussed in \cite{VorobevZyuzin2024}. A finite magnon thermal Hall effect in collinear and compensated N\'{e}el orders \cite{MatsumotoShindouMurakamiPRB2014} can exist only in weak ferromagnets and ferrimagnets \cite{Zyuzin2025b}.

In this Letter, we theoretically discuss the previously unexplored physics of magnon equilibrium spin currents in collinear antiferromagnets. Known examples of equilibrium currents in fermion systems include the supercurrent in superconductors \cite{Exp1,Exp2}, persistent currents in metallic rings in magnetic fields \cite{Kulik1970,SharvinSharvin1981,Exp3,Exp4}, and various equilibrium spin currents arising from spin-orbit coupling \cite{Rashba2003, Sonin2007, Tokatly2008, Sonin2010}. While supercurrents in superconductors and persistent currents are well understood both experimentally and theoretically, the existence of equilibrium spin currents in fermion systems remains largely an academic question \cite{Rashba2003, Sonin2007, Tokatly2008, Sonin2010}.
Magnon spin superfluidity and magnon spin supercurrents, on the other hand, are known to occur in magnon Bose-Einstein condensates in superfluid $^{3}$He-B \cite{Borovik-Romanov1984,Fomin1984,Borovik-Romanov1985,Borovik-Romanov1989,BunkovVolovik2010} and YIG \cite{Demokritov2006,BunkovPetrov2024}. These experiments require pumping of magnons, stabilization of a quasi-equilibrium magnon gas, and subsequent condensation of magnons into the lowest energy state.

In this Letter, we suggest that magnons in collinear antiferromagnets can exhibit equilibrium spin currents without the need for a magnon Bose-Einstein condensate. The presence of magnon equilibrium spin currents in the ground state of collinear antiferromagnets is a generic property. The only question is how to create such magnon equilibrium spin currents.
We show that a certain type of Dzyaloshinskii–Moriya interaction \cite{Dzyaloshinskii1958, Moriya1960b} can play the role of a static effective vector potential for magnons. Finite magnon equilibrium spin currents in response to the Dzyaloshinskii–Moriya interaction are thus expected. We then show that the Dzyaloshinskii–Moriya interaction can be controlled via an external electric field. This allows us to propose an Aharonov–Casher interference experiment in a ring geometry as a means to indirectly verify the existence of magnon equilibrium spin currents in collinear antiferromagnets.

\begin{figure}[t] 
\includegraphics[width=0.95 \columnwidth ]{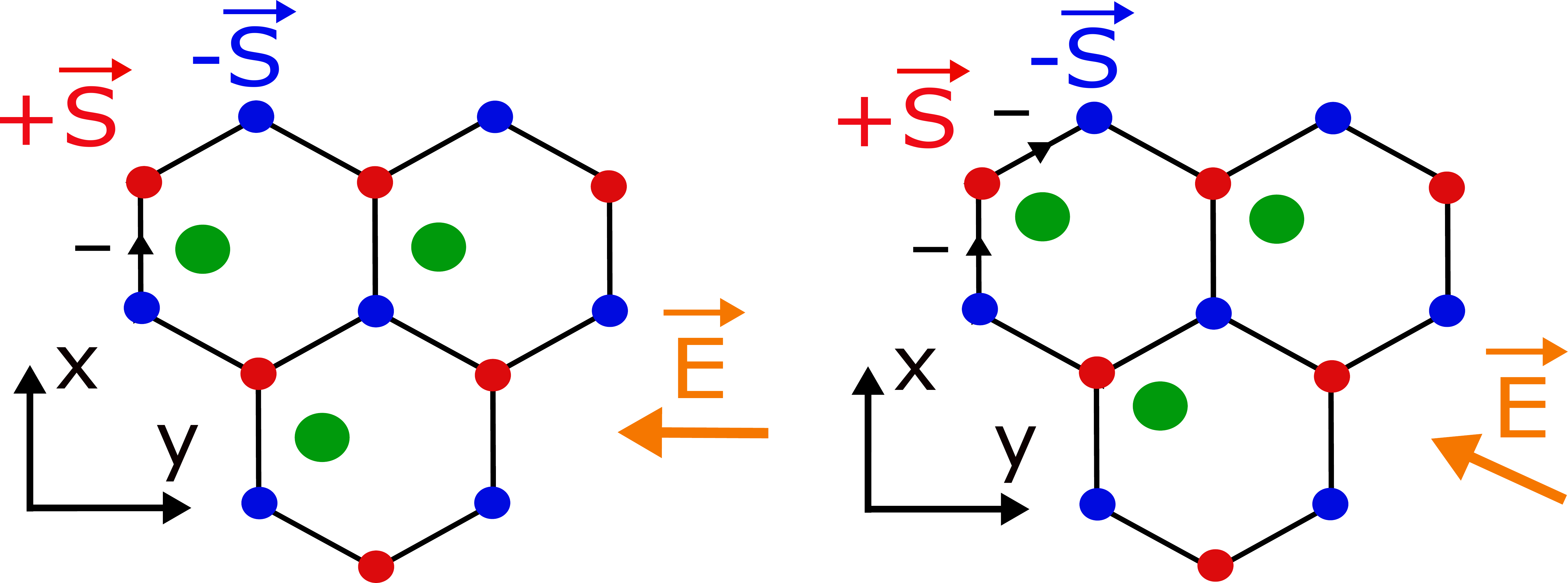} 

\protect\caption{Two collinear antiferromagnets on the honeycomb-like lattice. On the left is the genuine mirror-symmetric antiferromagnet while on the right is a ferrimagnet. Red and blue sites correspond to the N\'{e}el order with $\pm {\bf S}$ spins.  
The green atom is non-magnetic and its role is to create a Dzyaloshinskii-Moriya interaction on the links depicted here by the arrows. The $\pm$ signs correspond to the $\nu_{ij}=\pm$ in Eq. (\ref{exchange}). If the charges on the red/blue and green sites are of opposite signs, the external electric field ${\bf E}$ can tune the position of the green atom in the unit cell. }
\label{fig:fig1}  
\end{figure}

The antiferromagnetic system under study is shown in Fig. (\ref{fig:fig1}). The honeycomb-like lattice is chosen because it allows to demonstrate essential physics without any complications \cite{Zyuzin2025b}. 
Results presented below are general to any collinear antiferromagnet.
The spins ${\bf S}_{1}=+ {\bf S}$ and ${\bf S}_{2}=-{\bf S}$ are located at the red and blue sites correspondingly. 
For simplicity of the arguments we assume only the first nearest neighbor exchange interaction $J>0$ between localized spins. 
The green atom is non-magnetic and its role is to break the inversion symmetry and result in the Dzyaloshinskii-Moriya interaction between nearest-neighbor localized spins.
Overall Heisenberg exchange and Dzyaloshinskii-Moriya interaction between nearest-neighbor localized spins is
\begin{align}
H =& J\sum_{\langle ij \rangle} \left[\cos(\theta_{ij})\left(  S^{x}_{i}S^{x}_{j}  + S^{y}_{i}S^{y}_{j} \right) + S^{z}_{i}S^{z}_{j} \right]
\nonumber
\\
&
+J\sum_{\langle ij \rangle} \nu_{ij} \sin(\theta_{ij}) \left[ {\bf S}_{i}\times {\bf S}_{j} \right]_{z},
\label{exchange}
\end{align}
where $\theta$ is the parameter defining the Dzyaloshinskii-Moriya interaction. For example, $\theta_{ij} = \theta$ for the links with arrows in Fig. (\ref{fig:fig1}) and $\theta_{ij} = 0$ without, and $\nu_{ij}=\pm 1$ in accord with the arrows and signs on the links in Fig. (\ref{fig:fig1}). We observe that the N\'{e}el order is favorable in $z-$direction. It is worth emphasizing the origin of $\theta$. In Refs. \cite{Moriya1960b,Shekhtman1992,ZyuzinFietePRB2012,Zyuzin2025b} it was shown that if one considers a Hubbard model with spin-orbit coupling at half-filling, and integrates out the charge degrees of freedom, one will be left with the Heisenberg exchange and Dzyaloshinskii-Moriya interactions exactly in the form of Eq. (\ref{exchange}).
Parameter $\theta$ is defined by the nearest-neighbor fermion tunneling matrix elements $t$ and spin-orbit coupling $\lambda$ as $\theta = 2\arctan(\lambda/t)$. Heisenberg exchange interaction $J$ is affected by the spin-orbit coupling as $J = 4(t^2+\lambda^2)/U$, where $U$ is the on-site Hubbard repulsive interaction. Conventional Dzyaloshinskii-Moriya interaction constant is $D = J\sin(\theta)$.

We wish to study magnons that are fluctuations about the antiferromagnetic order.
We use Holstein-Primakoff transformation from spins to boson operators (for example, see \cite{ABP1967,Auerbach,Rezende}). 
For R (red) sublattice spin operators are $S_{i;\mathrm{R}}^{+} = S_{i;\mathrm{R}}^{x}+iS_{i;\mathrm{R}}^{y}=\sqrt{2S-a^{\dag}_{i}a_{i}}a_{i}$, $S_{i;\mathrm{R}}^{-}=a^{\dag}_{i}\sqrt{2S-a^{\dag}_{i}a_{i}}$  and $S_{\mathrm{R};i}^{z} = S-a_{i}^{\dag}a_{i}$, where $a_{i}^{\dag}$ and $a_{i}$ are boson operators and where $i$ denotes the unit cell number. For the B (blue) sublattice the spin operators are defined slightly differently, 
$S_{i;\mathrm{B}}^{+} =- S_{i;\mathrm{B}}^{x}+iS_{i;\mathrm{B}}^{y}=\sqrt{2S-b_{i}^{\dag}b_{i}}b_{i}^{\dag}$, $S_{i;\mathrm{B}}^{-}=b_{i}\sqrt{2S-b_{i}^{\dag}b_{i}}$  and $S_{i;\mathrm{B}}^{z} = -S+b_{i}^{\dag}b_{i}$, where $b_{i}^{\dag}$ and $b_{i}$ are boson operators. 
Fourier image of the Hamiltonian of magnons written in the basis $\hat{\Psi}^{\dag}_{\bf k} = (a_{\bf k}^{\dag}, ~ b_{\bf k}^{\dag}, ~ a_{-\bf k}, ~b_{-\bf k} ) $ is
\begin{align}\label{magnons}
\hat{H}_{\bf k} = 
SJ \left[
\begin{array}{cccc} 
3  & 0 & 0 & - \gamma_{\bf k}   \\
0 & 3   & - \gamma_{-\bf k}  & 0 \\
0 & - \gamma_{-\bf k}^{*} & 3   & 0 \\
- \gamma_{\bf k}^{*} & 0 & 0 & 3 
\end{array}
\right].
\end{align}
The Hamiltonian splits into two blocks described by $\hat{\Psi}^{\dag}_{{\bf k};\mathrm{I}} = (a_{\bf k}^{\dag} ,~b_{-\bf k} )$ and $\hat{\Psi}^{\dag}_{{\bf k};\mathrm{II}} = (b_{\bf k}^{\dag} ,~a_{-\bf k} )$ corresponding to two branches of the magnon modes \cite{ABP1967,Auerbach,Rezende}. The eigenvalue equation $\hat{H}_{\bf k}\psi_{\bf k} = \hat{\sigma}_{3}  \hat{E}_{\bf k}\psi_{\bf k}$ gives
\begin{align}\label{Ek}
\hat{E}_{\bf k} = \mathrm{diag}\left[\epsilon^{(\mathrm{I})}_{{\bf k};+},~ \epsilon^{(\mathrm{II})}_{{\bf k};+},~\epsilon^{(\mathrm{II})}_{{\bf k};-},~\epsilon^{(\mathrm{I})}_{{\bf k};-}\right],
\end{align}
where $\hat{\sigma}_{3} = \mathrm{diag}[1,1,-1,-1]$ in the basis of the Hamiltonian Eq. (\ref{magnons}) (also see \cite{commentA}). 
\begin{align}\label{spectrum}
\epsilon^{(\mathrm{I})}_{{\bf k};\pm} =
\pm SJ\sqrt{9 - \vert \gamma_{\bf k}\vert^2}, 
~~ \epsilon^{(\mathrm{II})}_{{\bf k};\pm} =  \pm SJ\sqrt{9 - \vert \gamma_{-\bf k}\vert^2},
\end{align}
corresponding to two branches of antiferromagnetic magnons \cite{ABP1967,Auerbach,Rezende} with different spin polarization.
The spectrum obeys a $\epsilon_{{\bf k};\pm}^{(\mathrm{I})} = -\epsilon_{-{\bf k};\mp}^{(\mathrm{II})}$ property, as expected \cite{MatsumotoShindouMurakamiPRB2014,ZyuzinKovalevPRL2016,Zyuzin2025b}. 
Spinors corresponding to the $\mathrm{I}$ block are 
\begin{align}\label{spinors}
\psi_{{\bf k};\mathrm{I};+} = \left[\begin{array}{c} \cosh\left( \frac{\xi_{\bf k}}{2}\right)e^{i\chi_{\bf k}} \\ \sinh\left( \frac{\xi_{\bf k}}{2}\right) \end{array}\right], 
\psi_{{\bf k};\mathrm{I};-} = \left[\begin{array}{c} \sinh\left( \frac{\xi_{\bf k}}{2}\right)e^{i\chi_{\bf k}} \\ \cosh\left( \frac{\xi_{\bf k}}{2}\right) \end{array}\right],
\end{align}
where $\cosh(\xi_{\bf k}) = 3/\sqrt{9-\vert \gamma_{\bf k}\vert^2}$, $\sinh(\xi_{\bf k}) = \vert \gamma_{\bf k}\vert/\sqrt{9-\vert \gamma_{\bf k}\vert^2}$ and $\gamma_{\bf k} = \vert \gamma_{\bf k}\vert e^{i\chi_{\bf k}}$. 
Spinors corresponding to the $\mathrm{II}$ block are $\psi_{{\bf k};\mathrm{II};\pm} = \psi_{-{\bf k};\mathrm{I};\pm} $. 
The spinors satisfy normalization $\psi_{{\bf k};\mathrm{I};\pm}^{\dag}\hat{\sigma}_{z}\psi_{{\bf k};\mathrm{I};\pm} = \hat{\sigma}_{z}$ and orthogonality $\psi_{{\bf k};\mathrm{I};\pm}^{\dag}\hat{\sigma}_{z}\psi_{{\bf k};\mathrm{I};\mp} = 0$ conditions.
The spin density is introduced as
\begin{align}\label{SpinOperator}
\hat{S}= \left[\begin{array}{cc} \hat{\tau}_{3} & 0 \\ 0 & \hat{\tau}_{3} \end{array}\right] \equiv \hat{\sigma}_{0}\hat{\tau}_{3}.
\end{align}
It can be checked that $\hat{S}\hat{\sigma}_{3}\hat{H} - \hat{H}\hat{\sigma}_{3}\hat{S} = 0$, therefore the spin operator is conserved in the system. 
The spin operator projected onto the $\mathrm{I}/\mathrm{II}$ blocks is
\begin{align}
\hat{S}_{\mathrm{I}/\mathrm{II}} = \pm \left[\begin{array}{cc}1 & 0 \\ 0 & -1 \end{array}\right] ,
\end{align}
therefore, $\psi_{{\bf k};\mathrm{I};\pm}^{\dag}\hat{S}_{\mathrm{I}}\psi_{{\bf k};\mathrm{I};\pm} = \pm 1 $ and $\psi_{{\bf k};\mathrm{II};\pm}^{\dag}\hat{S}_{\mathrm{II}}\psi_{{\bf k};\mathrm{II};\pm} = \mp 1 $ correspond to opposite spins. Magnon spin current corresponding to the magnon spin density operator is obtained from the continuity equation, and reads as
\begin{align}\label{CurrentDensity}
 {\bf J}^{\mathrm{s}}({\bf r}) = \hat{\Psi}^{\dag}({\bf r})\hat{S}\hat{\sigma}_{3} \hat{{\bf v}}\hat{\Psi}({\bf r}),
\end{align}
where velocity operator is defined as $\hat{{\bf v}} = i[\hat{H},{\bf r}]$, where ${\bf r}$ is the position operator.
Since we are interested in magnon equilibrium spin current, it is enough to pick diagonal part of the matrix element of the velocity operator. The off-diagonal parts are responsible for the magnon spin Nernst effect (see \cite{ZyuzinKovalevPRL2016} for details) and are traced to zero in the expression for the current in the equilibrium. 
In order to obtain total spin current, one has to estimate the current on the ground state of the system and integrate over the volume of the system, i.e. ${\bf j}^{\mathrm{s}} = \int_{\bf r}\langle {\bf J}^{\mathrm{s}}({\bf r}) \rangle$.
The Hilbert space of magnons is defined by the momentum ${\bf k}$.
The total spin current is then derived 
\begin{align}
j^{\mathrm{s}}_{\alpha} = \frac{1}{V} \sum_{{\bf k};n}(\hat{\sigma}_{3}\hat{\tau}_{3})_{nn}( \partial_{\alpha}\hat{E}_{\bf k} )_{nn}  g[ (\hat{E}_{\bf k})_{nn} ],
\end{align}
where $g(z) = (e^{z/T} - 1)^{-1}$ is the Bose-Einstein distribution function, and $T$ is the temperature. See supplemental material for details of the derivation of the spin current using three different approaches. 
Pauli matrices $\hat{\sigma}$ and $\hat{\tau}$ act on different spaces, and the index $n$ in, for example, $(\hat{\sigma}_{3}\hat{\tau}_{3})_{nn}$ stands for the elements of the $\hat{\sigma}_{3}\otimes\hat{\tau}_{3}$ matrix. 
We use $g(z)+1 = -g(-z)$ identity in obtaining the expression for the magnon equilibrium spin current,
\begin{align}\label{supercurrent}
j^{\mathrm{s}}_{\alpha} = \int_{\mathrm{BZ}} \frac{d{\bf k}}{(2\pi)^2} \left( \partial_{\alpha}\epsilon^{(\mathrm{I})}_{{\bf k};+} \right) \coth\left( \frac{\epsilon^{(\mathrm{I})}_{{\bf k};+}}{2T} \right) - (\mathrm{I} \rightarrow \mathrm{II}),
\end{align}
where integration is over the Brillouin zone ($\mathrm{BZ}$) and $(\mathrm{I} \rightarrow \mathrm{II})$ stands for an expression in which $\mathrm{I}$ block is substituted with $\mathrm{II}$. Therefore, if $ \gamma_{\bf k} =  \gamma_{-\bf k}$ then the equilibrium magnon spin current is zero.
It is known that Dzyaloshinskii-Moriya interaction can play the role of effective vector potential for magnons (for example \cite{Zyuzin2025b}) and it will make needed $ \gamma_{\bf k} \neq  \gamma_{-\bf k}$ inequality. 
Let us show that spin currents in equilibrium are indeed non-zero by the virtue of the Dzyaloshinskii-Moriya interaction.

\begin{figure}[t] 
\includegraphics[width=0.45 \columnwidth ]{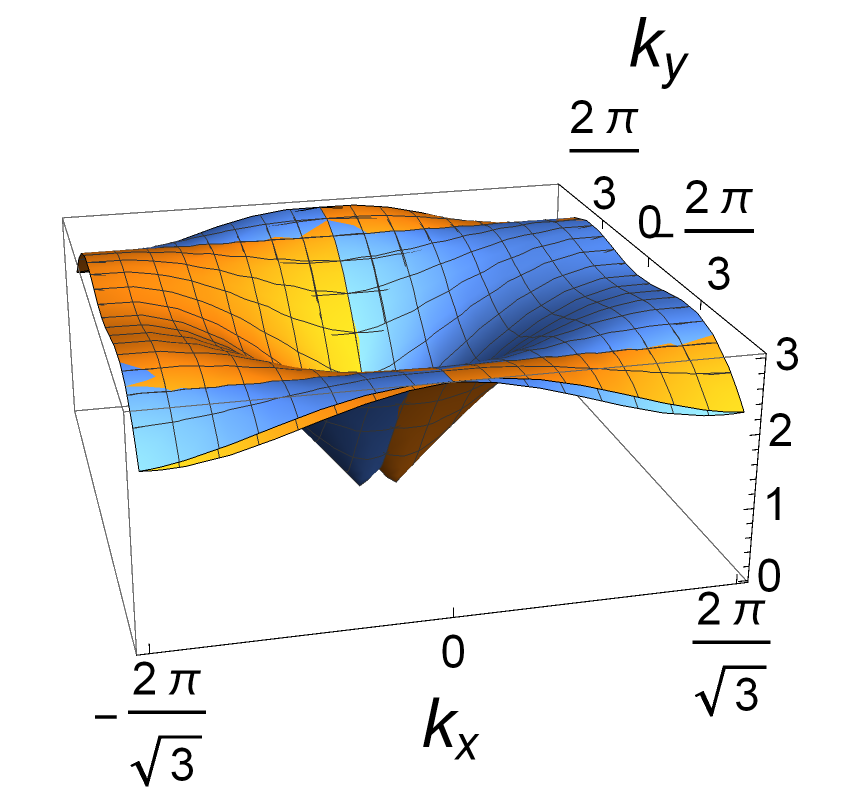}~~~ 
\includegraphics[width=0.45 \columnwidth ]{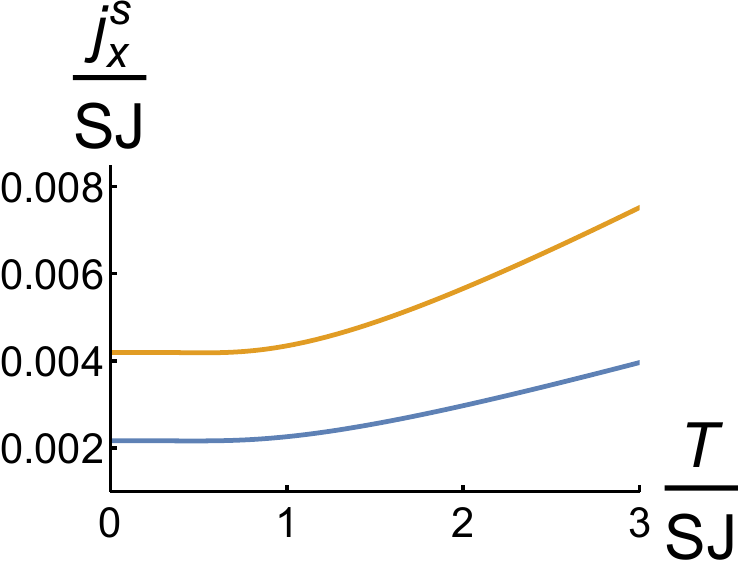} 

\protect\caption{Left: Spectrum of magnons for $\theta = 0.2$ of the genuine mirror-symmetric antiferromagnet. The plot emphasizes the spin-momentum splitting of magnons at the ${\bm \Gamma}$ point. Similar spin-momentum splitting of the magnon modes occurs for any position of the green atom in the unit cell. Right: Magnon equilibrium spin current in the genuine mirror-symmetric antiferromagnet plotted as a function of temperature for blue $\theta = -0.2$ and yellow $\theta = -0.4$. 
Parameter $SJ$ can reach $0.05 \mathrm{eV}$ in antiferromagnets. }
\label{fig:fig2}  
\end{figure}

\begin{figure}[t] 
\includegraphics[width=0.45 \columnwidth ]{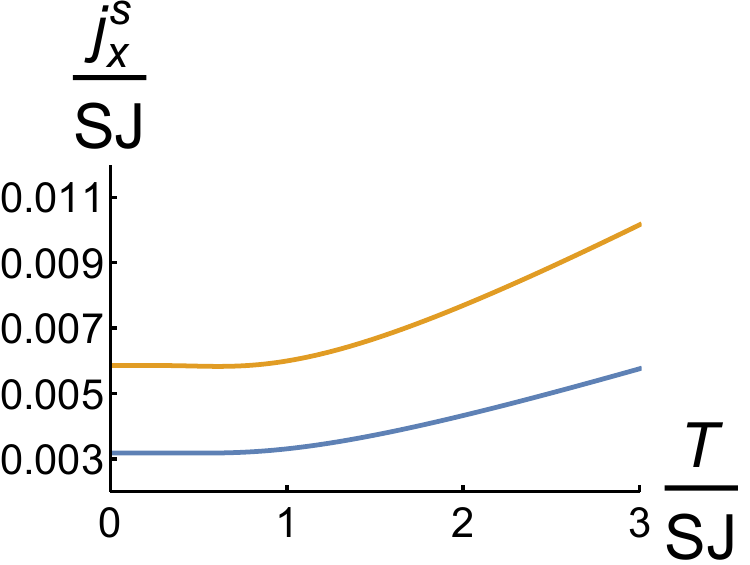} 
\includegraphics[width=0.45 \columnwidth ]{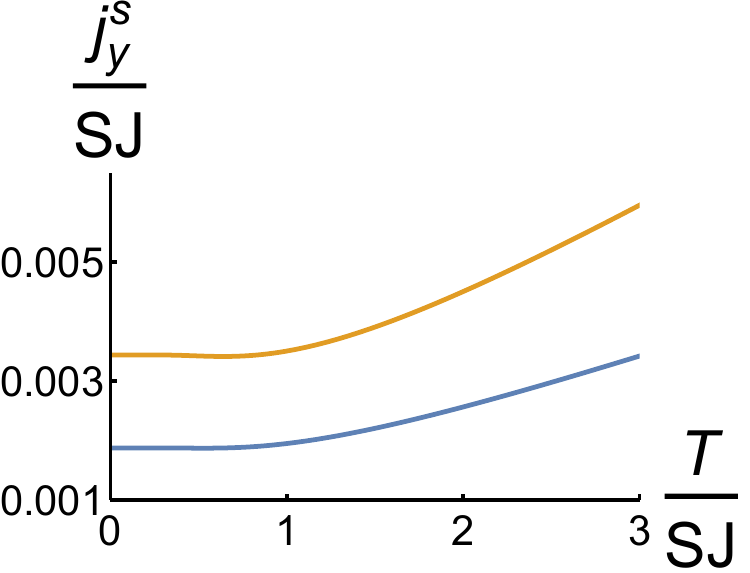} 

\protect\caption{Magnon equilibrium spin currents, $j_{x}^{\mathrm{s}}$ and $j_{y}^{\mathrm{s}}$ plotted as a function of temperature in the ferrimagnet phase for blue $\theta = -0.2$ and yellow $\theta =- 0.4$.  }
\label{fig:fig3}  
\end{figure}

\begin{figure}[h] 
\includegraphics[width=0.85 \columnwidth ]{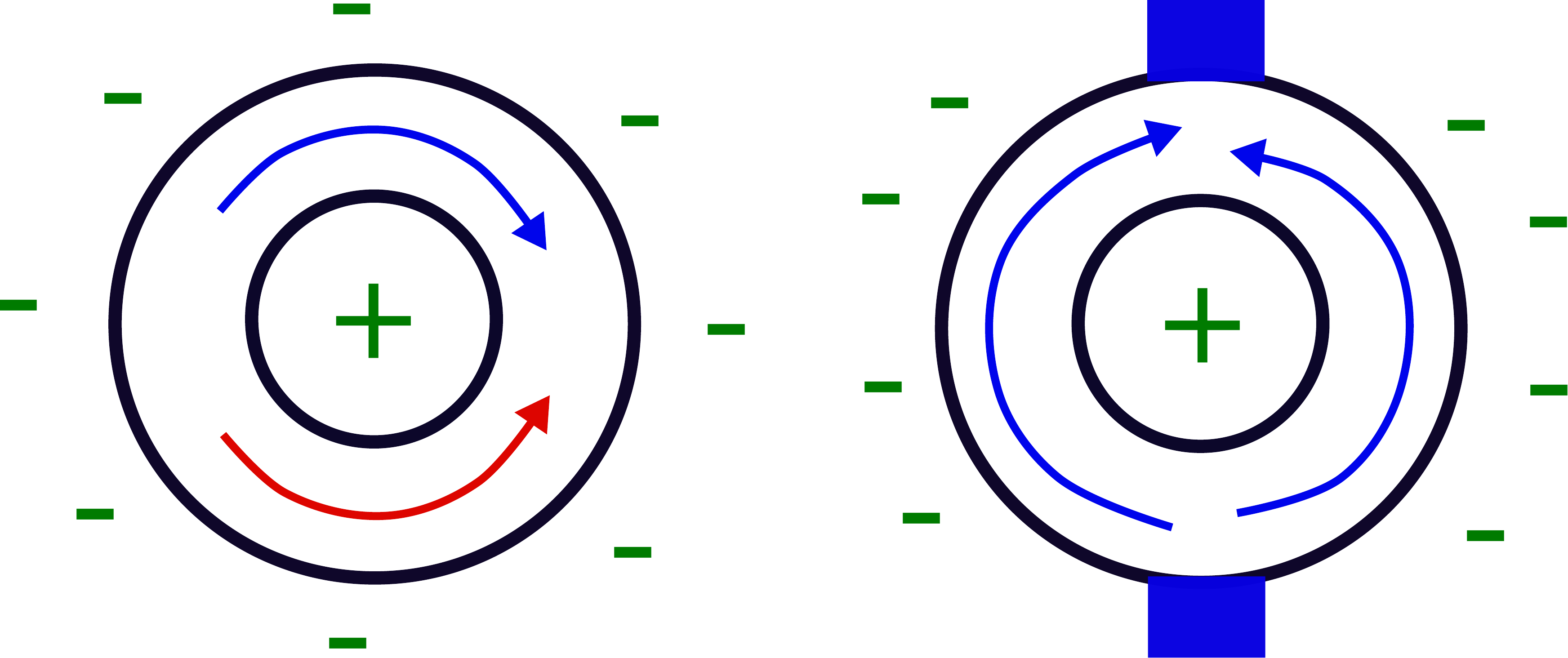} 

\protect\caption{Ring geometry. Static electric field (between + and -) creates a flux the magnons interact with. Left: there is a magnon equilibrium spin current as a result of the flux in the ring. Opposite spins denoted by red and blue propagate in opposite directions. Right: interference experiment. Blue rectangles are sources and sinks of the magnon spin polarized currents. Variation of the electric field would produce oscillations $\propto \cos(2\pi \theta R/a )$ in the interference of equal spin magnons propagated in the different sides of the ring. Here $R$ is the radius of the ring and $a$ is the distance between neighbor lattice sites. Recall that Dzyaloshinskii-Moriya interaction parameter $\theta$ is dimensionless and is linear in the electric field.}
\label{fig:fig4}  
\end{figure}

For the mirror-symmetric genuine antiferromagnet shown in the left of Fig. (\ref{fig:fig1}) we have the following nearest neighbor exchange interaction,
\begin{align}
&
\gamma_{\bf k} =  e^{i\frac{k_{x}}{2\sqrt{3}}} \left[2 \cos\left( \frac{k_{y}}{2} \right) +  e^{-i \left(\frac{\sqrt{3}}{2}k_{x} -\theta\right)} \right].
\end{align}
We observe that certain parts of the dispersion have a momentum shifted by the Dzyaloshinskii-Moriya interaction $\theta$.
The spin split spectrum of the two magnon modes is plotted in the left of Fig. (\ref{fig:fig2}). Magnon velocity is 
$\partial_{x}\epsilon^{(\mathrm{I}/\mathrm{II})}_{{\bf k};+} = -\left( \sqrt{3}/\sqrt{9- \vert \gamma_{\pm \bf k} \vert^2} \right) \cos\left( \frac{k_{y}}{2} \right)\sin\left( \frac{\sqrt{3}k_{x}}{2} \mp \theta\right)$. It is clear, that when $\theta = 0$, contributions from two blocks to Eq. (\ref{supercurrent}) cancel each other making the magnon equilibrium spin current exactly zero.
The magnon equilibrium spin current as a response to the Dzyaloshinskii-Moriya interaction $\theta$ is then ${\bf j}^{\mathrm{s}} = j^{\mathrm{s}}_{x}{\bf e}_{x}$. We  numerically plot the magnon equilibrium spin current as a function of temperature in the right of Fig. (\ref{fig:fig2}). 
Similar in structure magnon equilibrium spin currents will occur in the ferrimagnet phase \cite{Zyuzin2025b} as well. 
The ferrimagnet is shown in the right of Fig. (\ref{fig:fig1}). Nearest neighbor exchange interaction for the ferrimagnet phase reads as
\begin{align}
\gamma_{\bf k} &= 2 e^{\frac{i}{2}\left( \frac{k_{x}}{\sqrt{3}} - \theta \right) }\cos\left( \frac{k_{y}}{2} - \frac{\theta}{2}\right) + e^{-i\left( \frac{k_{x}}{\sqrt{3}} - \theta \right)}.
\end{align}
The magnon equilibrium spin current in the ferrimagnetic phase is then 
$
{\bf j}^{\mathrm{s}}  = j^{\mathrm{s}}_{x}  {\bf e}_{x} + j^{\mathrm{s}}_{y}  {\bf e}_{y} 
$. We plot both components of the current as a function of temperature in Fig. (\ref{fig:fig3}). 
We point out that the current in all of the plots, Fig. (\ref{fig:fig2}) and Fig. (\ref{fig:fig3}), is roughly linear in $\theta$ at zero temperature and becomes non-linear at temperatures $T \approx SJ$. The current doesn't vanish at zero temperature
and is almost temperature independent at small temperatures, i.e. at $T<SJ$. This is a consequence of the fact that the magnon equilibrium spin current is a property of the ground state rather than of the thermally enabled magnons.
Physically, it originates from the quantum correction (zero-point energy) to the ground state energy (see supplemental material for more details). We note that $SJ$ parameter may be large in antiferromagnets. Thus, the equilibrium magnon spin current can be experimentally relevant. We can imagine a situation when the green and red/blue atoms are oppositely charged. The total charge over the unit cell is zero, but there is a possibility of the finite dipole moment. When the green atom is at the center of the hexagon, the Dzyaloshinskii-Moriya interaction $\theta$ on the links is absent. By applying external electric field ${\bf E}$, one can tune the position of the green atom as, for example, shown in the Fig. (\ref{fig:fig1}). It is evident that the magnon equilibrium spin current is always perpendicular to the external electric field ${\bf E}$. Hence, the electric field plays the role of the effective magnetic field the magnons interact with. An example of such a field is shown in the left of Fig. (\ref{fig:fig4}).


Proposed in this Letter magnon equilibrium spin currents are analogous to diamagnetic currents in fermion systems. Thus, to probe these currents we propose an interference experiment schematically shown in Fig. (\ref{fig:fig4}) that should indirectly prove the existence of the magnon equilibrium spin current. There the blue square stands for a source and a sink of the spin polarized magnons. Then, equally spin-polarized magnons propagating in the different shoulders of the ring (shown by blue arrows) will acquire different phases $\propto e^{\pm i \pi \theta R/a}$, where $R$ is the radius of the ring and $a$ is the distance between neighbor lattice sites and $\theta$ is the Dzyaloshinskii-Moriya interaction dimensionless parameter. Therefore, their interference at the sink will result in $\cos(2\pi \theta R/a )$ oscillation of the amplitude of the wave-function of the propagated magnons. This is essentially the Aharonov-Casher effect \cite{AharonovCasher} for magnons. We note that the Aharonov-Casher effect in the magnon Bose-Einstein condensates occurring in YIG has been theoretically discussed in \cite{Kouki2014}. 
Furthermore, to directly prove the existence of quantum contributions to the magnon spin equilibrium current, we propose to study the amplitude of oscillations as a function of temperature. According to our theory, it shouldn't drastically change as a function of temperature at low temperatures.

We leave a question of possible conversion of the nonequilibrium magnon spin current to the magnon equilibrium spin current, just like what occurs in superconductors, for future research. The nonequilibrium spin currents in collinear antiferromagnets are either dissipative anisotropic magnon spin currents (spin-splitter) \cite{Naka2019,Exp2026} or magnon spin Nernst effect \cite{ZyuzinKovalevPRL2016} or magnon thermal Hall effect \cite{MatsumotoShindouMurakamiPRB2014,Zyuzin2025b}. For example, in the genuine antiferromagnet (left Fig. (\ref{fig:fig1})) phase the spin-splitter magnon current is
${\bf j}^{\Gamma} = \sigma_{xy}\left[ (\partial_{y}\chi){\bf e}_{x} + (\partial_{x}\chi){\bf e}_{y} \right]$, while in ferrimagnet (right Fig. (\ref{fig:fig1})) phase ${\bf j}^{\Gamma} = \sigma_{\parallel} \left[ (\partial_{y}\chi){\bf e}_{y} - (\partial_{x}\chi){\bf e}_{x} \right]$, where ${\bm \partial}\chi$ is the applied temperature gradient. On the other hand, magnon spin Nernst effect is ${\bf j}^{\mathrm{MSNE}} = \sigma_{\mathrm{MSNE}}\left[ (\partial_{y}\chi){\bf e}_{x} - (\partial_{x}\chi){\bf e}_{y} \right]$.
It can be seen from the structure of the currents that nonequilibrium magnon spin currents and predicted in this Letter magnon equilibrium spin current don't share anything in common as the latter is a response to effective magnon vector potential.
Furthermore, it is interesting to explore the possibility of a reduction of the electric field and equivalently of the parameter $\theta$ as a result of the magnon equilibrium spin current. This would be the analog of the magnetic field suppression in the bulk of the superconductor of the first type.

We note that magnons in insulating ferromagnets will also show predicted here magnon equilibrium spin current due to the Dzyaloshinskii-Moriya interaction. However, temperature dependence of the currents is expected to start from zero at zero temperature and have a similar dependence as the ones shown in the right of Fig. (\ref{fig:fig2}) and Fig. (\ref{fig:fig3}). This is because $\coth(z/2T)$ in the expression for the magnon equilibrium spin current in collinear antiferromagnets Eq. (\ref{supercurrent}) in case of ferromagnetic magnons will be replaced by the Bose-Einstein distribution function $g(z)$. Thus, magnon equilibrium spin currents in ferromagnets are analogs of persistent currents in normal metals \cite{Exp3,Exp4}, while magnon equilibrium spin currents in collinear antiferromagnets are analogs of supercurrents in superconductors \cite{Exp1,Exp2}.

To conclude, we have theoretically uncovered a previously unexplored magnon equilibrium spin currents in collinear antiferromagnets that flow due to the Dzyaloshinskii-Moriya interaction \cite{Dzyaloshinskii1958, Moriya1960b}. Such a spin current being a response to the effective magnon vector potential is the magnon analog of supercurrent in superconductors \cite{Exp1,Exp2}. Spin supercurrents of magnons were known in the context of Bose-Einstein condensates of magnons in $^{3}$He-B and YIG. However, as we showed in this Letter, collinear antiferromagnets can also show the spin supercurrents in their ground state and there is no need for the Bose-Einstein condensate of magnons for these currents to exist. We have proposed an interference experiment to explore magnon equilibrium spin currents through the Aharonov-Casher effect \cite{AharonovCasher}.

The author is grateful to Pirinem School of Theoretical Physics for hospitality. This work is supported by FFWR-2024-0016.


\begin{widetext}
\newpage
\setcounter{equation}{0}

\section{Derivation of the equilibrium spin current}

\subsection{Continuity equation}
It is instructive to first derive the spin current using continuity equation. 
We introduce a general operator 
\begin{align}
\hat{O}({\bf r}) = \hat{\Psi}^{\dag}({\bf r})\hat{O}\Psi({\bf r}),
\end{align}
where $\hat{\Psi}^{\dag}({\bf r}) = [\nu_{1}({\bf r}),..,\nu_{N}({\bf r}),\nu_{1}^{\dag}({\bf r}),..,\nu_{N}^{\dag}({\bf r})]$ are the boson operators describing the extended Gor'kov-Nambu space of magnons, $[\nu_{i}({\bf r}),\nu_{j}^{\dag}({\bf r}^{\prime})]= \delta_{i,j}\delta_{{\bf r},{\bf r}^{\prime}}$. 
\begin{align}
&
[\Psi_{i}({\bf r}),\Psi_{j}^{\dag}({\bf r})] = (\sigma_{3})_{ij}\delta_{{\bf r},{\bf r}^\prime},
\\
&
[\Psi_{i}^{\dag}({\bf r}),\Psi_{j}^{\dag}({\bf r})] = -i (\sigma_{2})_{ij}\delta_{{\bf r},{\bf r}^\prime},
\\
&
[\Psi_{i}({\bf r}),\Psi_{j}({\bf r})] = i (\sigma_{2})_{ij}\delta_{{\bf r},{\bf r}^\prime}.
\end{align}

Then, the time derivative of this operator reads 
\begin{align}
\frac{\partial \hat{O}({\bf r})}{\partial t} = i \left[ \hat{H},\hat{O}({\bf r}) \right] ,
\end{align} 
where 
\begin{align}
\hat{H} = \frac{1}{2} \int d{\bf r}^{\prime} \hat{\Psi}^{\dag}({\bf r}^{\prime})\hat{H}\Psi({\bf r}^{\prime})
\end{align} 
is the Hamiltonian. The Hamiltonian acts on the function as
\begin{align}
\hat{H}\hat{\Psi}({\bf r}) = \sum_{\{{\bm \delta} \}}H_{\delta}\hat{\Psi}({\bf r}+{\bm \delta}),
\end{align}
where $\delta$ is the vector connecting neighboring sites on the lattice.
\begin{align}
\hat{\Psi}^{\dag}({\bf r})\hat{O}\Psi({\bf r}) \hat{\Psi}^{\dag}({\bf r}^{\prime})\hat{H}\Psi({\bf r}^{\prime})
&
=
\hat{O}_{nn^{\prime}} (\hat{H})_{mm^{\prime}}\hat{\Psi}^{\dag}_{n}({\bf r})\Psi_{n^\prime}({\bf r}) \hat{\Psi}^{\dag}_{m}({\bf r}^{\prime})
\Psi_{m^{\prime}}({\bf r}^{\prime})
\\
&
=
\sum_{\{{\bm \delta} \}}\hat{O}_{nn^{\prime}} (\hat{H}_{{\bm \delta}})_{mm^{\prime}}\hat{\Psi}^{\dag}_{n}({\bf r})\Psi_{n^\prime}({\bf r}) \hat{\Psi}^{\dag}_{m}({\bf r}^{\prime})
\Psi_{m^{\prime}}({\bf r}^{\prime}+{\bm \delta}).
\end{align}
To proceed further we use
\begin{align}
&
\hat{\Psi}_{n^\prime}({\bf r}) \hat{\Psi}^{\dag}_{m}({\bf r}^{\prime}) =  \hat{\Psi}^{\dag}_{m}({\bf r}^{\prime}) \hat{\Psi}_{n^\prime}({\bf r}) + (\sigma_{3})_{n^\prime,m} \delta_{{\bf r},{\bf r}^{\prime}},
\\
&
\hat{\Psi}_{n}^{\dag}({\bf r}) \hat{\Psi}^{\dag}_{m}({\bf r}^{\prime})
=\hat{\Psi}^{\dag}_{m}({\bf r}^{\prime})\hat{\Psi}_{n}^{\dag}({\bf r}) - i(\sigma_{2})_{n,m} \delta_{{\bf r},{\bf r}^{\prime}},
\\
&
\hat{\Psi}_{n^\prime}({\bf r}) \hat{\Psi}_{m^\prime}({\bf r}^{\prime}+{\bm \delta})
=\hat{\Psi}_{m^\prime}({\bf r}^{\prime}+{\bm \delta})\hat{\Psi}_{n^\prime}({\bf r}) + i(\sigma_{2})_{n^\prime,m^\prime} \delta_{{\bf r},{\bf r}^{\prime}+{\bm \delta}},
\\
&
\hat{\Psi}_{n}^{\dag}({\bf r}) \hat{\Psi}_{m^\prime}({\bf r}^{\prime}+{\bm \delta}) = 
\hat{\Psi}_{m^\prime}({\bf r}^{\prime}+{\bm \delta})\hat{\Psi}_{n}^{\dag}({\bf r}) - (\sigma_{3})_{n,m} \delta_{{\bf r},{\bf r}^{\prime}+{\bm \delta}}.
\end{align} 
Then 
\begin{align}
\hat{\Psi}^{\dag}_{n}({\bf r})\Psi_{n^\prime}({\bf r}) \hat{\Psi}^{\dag}_{m}({\bf r}^{\prime})
\Psi_{m^{\prime}}({\bf r}^{\prime}+{\bm \delta}) 
&
= 
\hat{\Psi}^{\dag}_{m}({\bf r}^{\prime})
\Psi_{m^{\prime}}({\bf r}^{\prime}+{\bm \delta}) 
\hat{\Psi}^{\dag}_{n}({\bf r})\Psi_{n^\prime}({\bf r})
\\
&
+\hat{\Psi}^{\dag}_{n}({\bf r}) (\sigma_{3})_{n^\prime m} \delta_{{\bf r},{\bf r}^{\prime}}\Psi_{m^{\prime}}({\bf r}^{\prime}+{\bm \delta}) 
\\
&
+\hat{\Psi}^{\dag}_{m}({\bf r}^{\prime}) i(\sigma_{2})_{n^\prime m^\prime} \delta_{{\bf r},{\bf r}^{\prime}+{\bm \delta}} \hat{\Psi}^{\dag}_{n}({\bf r})
\\
&
-\hat{\Psi}_{n^\prime}({\bf r}) i(\sigma_{2})_{n m} \delta_{{\bf r},{\bf r}^{\prime}} \hat{\Psi}_{m^\prime}({\bf r}^\prime + {\bm \delta})
\\
&
-\hat{\Psi}^{\dag}_{m}({\bf r}^\prime) (\sigma_{3})_{m^\prime n} \delta_{{\bf r},{\bf r}^{\prime}+{\bm \delta}}\Psi_{n^{\prime}}({\bf r}) .
\end{align}
An identity $\hat{\Psi}^{\dag} \sigma_{1}= \hat{\Psi}^{\mathrm{T}}$ (that is the symplectic relation in the Bogoliubov-de Gennes theory) as well as $\sigma_{1} H_{{\bm \delta}}\sigma_{1} = H_{-{\bm \delta}}^{\mathrm{T}}$ (where $\mathrm{T}$ is the transposition of the matrix structure only) will be used to perform the following transformations. 
The fifth term in the expression above is transformed as
\begin{align}
&
- \int d{\bf r}^{\prime} \hat{O}_{nn^{\prime}} (\hat{H}_{{\bm \delta}})_{mm^{\prime}}
\hat{\Psi}^{\dag}_{m}({\bf r}^\prime) (\sigma_{3})_{m^\prime n} \delta_{{\bf r},{\bf r}^{\prime}+{\bm \delta}}\Psi_{n^{\prime}}({\bf r}) 
 =
 -
 \hat{\Psi}^{\dag}_{m}({\bf r}-{\bm \delta})   (\hat{H}_{{\bm \delta}})_{mm^{\prime}}(\sigma_{3})_{m^\prime n} \hat{O}_{nn^{\prime}} \Psi_{n^{\prime}}({\bf r}) ,
\end{align}
the third term is
\begin{align}
 \int d{\bf r}^{\prime} \hat{O}_{nn^{\prime}} (\hat{H}_{{\bm \delta}})_{mm^{\prime}}
\hat{\Psi}^{\dag}_{m}({\bf r}^{\prime}) i(\sigma_{2})_{n^\prime m^\prime} \delta_{{\bf r},{\bf r}^{\prime}+{\bm \delta}} \hat{\Psi}^{\dag}_{n}({\bf r})
&
 =
 \Psi_{n}^{\dag}({\bf r}) 
 \hat{O}_{nn^{\prime}} (\sigma_{1}\hat{H}_{-{\bm \delta}}^{\mathrm{T}} \sigma_{1})_{mm^{\prime}} 
 \hat{\Psi}^{\dag}_{m}({\bf r}-{\bm \delta})  i(\sigma_{2})_{n^\prime m^\prime}
\\
 &
 =
\Psi_{n}^{\dag}({\bf r})   \hat{O}_{nn^{\prime}} (\sigma_{3})_{n^\prime m^\prime} (\hat{H}_{{\bm \delta}})_{m^{\prime}m}  \hat{\Psi}_{m}({\bf r}+{\bm \delta}),
\end{align}
where we have used the fact that $ \hat{\Psi}^{\dag}_{m}({\bf r}-{\bm \delta}) \Psi_{n}^{\dag}({\bf r})  = \Psi_{n}^{\dag}({\bf r}) \hat{\Psi}^{\dag}_{m}({\bf r}-{\bm \delta})$ because ${\bm \delta} \neq 0$ and the two points never coincide.
Possible terms with ${\bm \delta} = 0$ in the Hamiltonian don't contribute to the current. We used the symplectic relation and relabeled ${\bm \delta}$ to $-{\bm \delta}$ in the sum over ${\bm \delta}$.

\begin{align}
 \int d{\bf r}^{\prime} \hat{O}_{nn^{\prime}} (\hat{H}_{{\bm \delta}})_{mm^{\prime}}
\hat{\Psi}^{\dag}_{n}({\bf r}) (\sigma_{3})_{n^\prime m} \delta_{{\bf r},{\bf r}^{\prime}}\Psi_{m^{\prime}}({\bf r}^{\prime}+{\bm \delta}) =
 \hat{\Psi}^{\dag}_{n}({\bf r})\hat{O}_{nn^{\prime}}  (\sigma_{3})_{n^\prime m} (\hat{H}_{{\bm \delta}})_{mm^{\prime}}\Psi_{m^{\prime}}({\bf r}+{\bm \delta}) 
\end{align}

\begin{align}
-
 \int d{\bf r}^{\prime} \hat{O}_{nn^{\prime}} (\hat{H}_{{\bm \delta}})_{mm^{\prime}}\hat{\Psi}_{n^\prime}({\bf r}) i(\sigma_{2})_{n m} \delta_{{\bf r},{\bf r}^{\prime}} \hat{\Psi}_{m^\prime}({\bf r}^\prime + {\bm \delta})
\end{align}

Velocity operator reads as 
\begin{align}
{\bf v} = i \sum_{{\bm \delta}} e^{i{\bm \delta}\hat{{\bf p}}} {\bm \delta} \hat{H}_{{\bm \delta}},
\end{align}
where $\hat{{\bf p}} = - i {\bm \partial}$ is the momentum operator, such that  $e^{i{\bm \delta}\hat{{\bf p}}}$ is just the translation operator that shifts the coordinate in ${\bm \delta}$.
Then we can write for the current
\begin{align}
&
i \sum_{{\bm \delta}}
\left[ 
- \Psi^{\dag}({\bf r}-{\bm \delta}) \hat{H}_{{\bm \delta}}\sigma_{3} \hat{O} \Psi({\bf r}) 
+  \Psi^{\dag}({\bf r}) \hat{O} \sigma_{3} \hat{H}_{{\bm \delta}} \Psi({\bf r}+{\bm \delta})    \right]
\\
&
=
i \sum_{{\bm \delta}}
\left[ 
- \Psi^{\dag}({\bf r}-{\bm \delta}) \hat{H}_{{\bm \delta}}\sigma_{3} \hat{O} \Psi({\bf r}-{\bm \delta}+{\bm \delta}) 
+  \Psi^{\dag}({\bf r}+{\bm \delta}-{\bm \delta}) \hat{O} \sigma_{3} \hat{H}_{{\bm \delta}} \Psi({\bf r}+{\bm \delta})    \right]
\\
&
=
i \sum_{{\bm \delta}}
\left[ 
- \Psi^{\dag}({\bf r}-{\bm \delta}) \hat{H}_{{\bm \delta}}\sigma_{3} \hat{O} e^{i{\bm \delta}\hat{{\bf p}}}\Psi({\bf r}-{\bm \delta}) 
+  \Psi^{\dag}({\bf r}+{\bm \delta}) e^{i{\bm \delta}\hat{{\bf p}}} \hat{O} \sigma_{3} \hat{H}_{{\bm \delta}} \Psi({\bf r}+{\bm \delta})    \right],
\end{align}
then 
$\Psi^{\dag}({\bf r} \pm {\bm \delta}) \approx \Psi^{\dag}({\bf r})  \pm {\bm \delta}\cdot {\partial}\Psi^{\dag}({\bf r})  $ and the same for $\Psi({\bf r} \pm {\bm \delta})$ fields. Then
\begin{align}
&
i \sum_{{\bm \delta}}
\left[ 
- \Psi^{\dag}({\bf r}-{\bm \delta}) \hat{H}_{{\bm \delta}}\sigma_{3} \hat{O} e^{i{\bm \delta}\hat{{\bf p}}}\Psi({\bf r}-{\bm \delta}) 
+  \Psi^{\dag}({\bf r}+{\bm \delta}) e^{i{\bm \delta}\hat{{\bf p}}} \hat{O} \sigma_{3} \hat{H}_{{\bm \delta}} \Psi({\bf r}+{\bm \delta})    \right]
\\
&
=
i \Psi^{\dag}({\bf r}) \left[  \hat{H}\sigma_{3} \hat{O} - \hat{O} \sigma_{3} \hat{H} \right]\Psi({\bf r})
-
{\bm \partial}\cdot \left[ \Psi^{\dag}({\bf r}) \left( {\bf v} \sigma_{3} \hat{O} + \hat{O} \sigma_{3} {\bf v}  \right)\Psi({\bf r}) \right],
\end{align}
therefore, for the current of the $\hat{O}$ operator to be a well defined quantity, one has to require
\begin{align}
 \hat{H}\sigma_{3} \hat{O} - \hat{O} \sigma_{3} \hat{H} = 0.
\end{align} 
The spin current will be defined for $\hat{O} = \hat{S}$ defined in the Main Text,
\begin{align}
{\bf j}_{\mathrm{s}} = \Psi^{\dag}({\bf r}) \left( {\bf v} \sigma_{3} \hat{S} + \hat{S} \sigma_{3} {\bf v}  \right)\Psi({\bf r})
\end{align}

\subsection{Calculating spin current}
In the main text we have evaluated expectation value of the magnon equilibrium spin current Eq. (10) of the Main Text using rules outlined in Ref. \cite{ZyuzinKovalevPRL2016}) (as well as in Ref. \cite{MatsumotoShindouMurakamiPRB2014}). Here we show a more detailed derivation of Eq. (10) of the Main Text by explicitly showing the steps. We will follow the lines of Ref. \cite{ZyuzinKovalevPRL2016}. 
The diagonal basis for the $\mathrm{I}/\mathrm{II}$ block defined by $\hat{\Psi}_{{\bf k};\mathrm{I}}^{\dag} = (a_{\bf k}^{\dag} ,~b_{-{\bf k}}) $ and $\hat{\Psi}_{{\bf k};\mathrm{II}}^{\dag} = (b_{\bf k}^{\dag} ,~a_{-{\bf k}}) $ is
\begin{align}
&
\hat{\Psi}_{{\bf k};\mathrm{I}} = \hat{T}_{{\bf k};\mathrm{I}}\hat{\Gamma}_{{\bf k};\mathrm{I}},~~
\hat{\Psi}_{{\bf k};\mathrm{I}}^{\dag} = \hat{\Gamma}_{{\bf k};\mathrm{I}}^{\dag}\hat{T}_{{\bf k};\mathrm{I}}^{\dag},
\\
&
\hat{\Psi}_{{\bf k};\mathrm{II}} = \hat{T}_{{\bf k};\mathrm{II}}\hat{\Gamma}_{{\bf k};\mathrm{II}},~~
\hat{\Psi}_{{\bf k};\mathrm{II}}^{\dag} = \hat{\Gamma}_{{\bf k};\mathrm{II}}^{\dag}\hat{T}_{{\bf k};\mathrm{II}}^{\dag},
\end{align}
where 
$\hat{\Gamma}_{{\bf k};\mathrm{I}}^{\dag} = (c^{\dag}_{{\bf k};1}, ~c_{-{\bf k};2})$, $\hat{\Gamma}_{{\bf k};\mathrm{II}}^{\dag} = (c^{\dag}_{{\bf k};2}, ~c_{-{\bf k};1})$ are normal modes,
and the paraunitary matrix that diagonalizes the Hamiltonian is made of the spinors Eq. (5) of the Main Text, i.e.
\begin{align}
\hat{T}_{{\bf k};\mathrm{I}/\mathrm{II}} =\left[ \begin{array}{cc} \cosh\left( \frac{\xi_{\bf k}}{2}\right)e^{\pm i\chi_{\bf k}}  & \sinh\left( \frac{\xi_{\bf k}}{2}\right)e^{\pm i\chi_{\bf k}} \\  \sinh\left( \frac{\xi_{\bf k}}{2}\right) & \cosh\left( \frac{\xi_{\bf k}}{2}\right) \end{array}\right],
\end{align}
where $\xi_{\bf k}$ and $\chi_{\bf k}$ are defined in Eq. (5) of the Main Text, and the matrix obeys $\hat{T}_{{\bf k};\mathrm{I}/\mathrm{II}}^{\dag} \hat{\sigma}_{3} \hat{T}_{{\bf k};\mathrm{I}/\mathrm{II}} =\hat{T}_{{\bf k};\mathrm{I}/\mathrm{II}} \hat{\sigma}_{3} \hat{T}_{{\bf k};\mathrm{I}/\mathrm{II}}^{\dag} = \hat{\sigma}_{3}$.
In this way
\begin{align}
\hat{\Psi}_{{\bf k};\mathrm{I}}^{\dag}\hat{H}_{{\bf k};\mathrm{I}}\hat{\Psi}_{{\bf k};\mathrm{I}}
&
 =\hat{\Gamma}_{{\bf k};\mathrm{I}}^{\dag}( \hat{\sigma}_{3}\hat{E}_{\bf k} )^{(\mathrm{I})}\hat{\Gamma}_{{\bf k};\mathrm{I}}
 \nonumber
\\
&
 =\hat{\Gamma}_{{\bf k};\mathrm{I}}^{\dag} \mathrm{diag}[\epsilon^{(\mathrm{I})}_{{\bf k};+},~ -\epsilon^{(\mathrm{I})}_{{\bf k};-}]\hat{\Gamma}_{{\bf k};\mathrm{I}}
  \nonumber
 \\
 &
 =\epsilon^{(\mathrm{I})}_{{\bf k};+} c^{\dag}_{{\bf k};1} c_{{\bf k};1} 
 -\epsilon^{(\mathrm{I})}_{{\bf k};-} c_{-{\bf k};2}c_{-{\bf k};2}^{\dag},
\end{align}
where $\hat{E}_{\bf k}$ is defined in Eq. (3) of the Main Text, and where $\mathrm{diag}[..,~..]$ is a diagonal matrix.
For the second block 
\begin{align}
\hat{\Psi}_{{\bf k};\mathrm{II}}^{\dag}\hat{H}_{{\bf k};\mathrm{II}}\hat{\Psi}_{{\bf k};\mathrm{II}}
&
 =\hat{\Gamma}_{{\bf k};\mathrm{II}}^{\dag}( \hat{\sigma}_{3}\hat{E}_{\bf k} )^{(\mathrm{II})}\hat{\Gamma}_{{\bf k};\mathrm{II}}
 \nonumber
\\
&
 =\hat{\Gamma}_{{\bf k};\mathrm{II}}^{\dag} \mathrm{diag}[\epsilon^{(\mathrm{II})}_{{\bf k};+},~ -\epsilon^{(\mathrm{II})}_{{\bf k};-}]\hat{\Gamma}_{{\bf k};\mathrm{II}}
 \nonumber
 \\
 &
 =\epsilon^{(\mathrm{II})}_{{\bf k};+} c^{\dag}_{{\bf k};2} c_{{\bf k};2} 
 -\epsilon^{(\mathrm{II})}_{{\bf k};-} c_{-{\bf k};1}c_{-{\bf k};1}^{\dag}.
 \label{diagonal}
\end{align}
We can read off a relation $\epsilon^{(\mathrm{I}/\mathrm{II})}_{-{\bf k};-} = - \epsilon^{(\mathrm{II}/\mathrm{I})}_{{\bf k};+}$, as expected. In addition, in our particular studied system, there is a $\epsilon^{(\mathrm{I}/\mathrm{II})}_{{\bf k};+} = - \epsilon^{(\mathrm{I}/\mathrm{II})}_{{\bf k};-}$ relation (see Eq. (3) of the Main Text).
Therefore, expectation values are 
\begin{align}
&
\langle  c^{\dag}_{{\bf k};1} c_{{\bf k};1}  \rangle = g(\epsilon^{(\mathrm{I})}_{{\bf k};+}),
~~~
\langle c_{{\bf k};2}^{\dag}c_{{\bf k};2} \rangle = g(\epsilon^{(\mathrm{II})}_{{\bf k};+}) ,
\end{align}
where $g(z) = (e^{z/T}-1)^{-1}$ is the Bose-Einstein distribution function. 
Therefore,
\begin{align}
\langle c_{-{\bf k};2}c_{-{\bf k};2}^{\dag} \rangle 
&= 1+ g(\epsilon^{(\mathrm{II})}_{-{\bf k};+}) 
= 1+ g(-\epsilon^{(\mathrm{I})}_{{\bf k};-})
 \nonumber
\\
&
 =- g(\epsilon^{(\mathrm{I})}_{{\bf k};-})
=1+g(\epsilon^{(\mathrm{I})}_{{\bf k};+}),
\label{correlator1}
\end{align}
where we have used boson commutation relations, $g(-z) = -1-g(z)$, and the two relations between spectrum branches discussed after Eq. (\ref{diagonal}). Another expectation value is 
\begin{align}
\label{correlator2}
\langle c_{-{\bf k};1}c_{-{\bf k};1}^{\dag} \rangle 
=1+g(\epsilon^{(\mathrm{II})}_{{\bf k};+}).
\end{align}
The spin operator was derived in Ref. \cite{ZyuzinKovalevPRL2016} and is given in Eq. (6) of the Main Text. 
The spin current density is derived from the continuity equation in accord with Ref. \cite{ZyuzinKovalevPRL2016}, and is given in Eq. (8) of the Main Text. 
Projections of the $\hat{S} \hat{\sigma}_{3}$ operator on the $\mathrm{I}/\mathrm{II}$ blocks are $(\hat{S} \hat{\sigma}_{3})^{(\mathrm{I}/\mathrm{II})} = \pm 1 $.
Therefore, contribution of the $\mathrm{I}/\mathrm{II}$ block to the spin current density reads
\begin{align}
\left( j_{{\bf k};\alpha}^{\mathrm{s}}\right)^{(\mathrm{I}/\mathrm{II})} 
&
= \langle \hat{\Psi}_{{\bf k};\mathrm{I}/\mathrm{II}}^{\dag} (\hat{S} \hat{\sigma}_{3})^{(\mathrm{I}/\mathrm{II})} \hat{v}^{(\mathrm{I}/\mathrm{II})}_{\alpha} \hat{\Psi}_{{\bf k};\mathrm{I}/\mathrm{II}}  \rangle
\\
&
\rightarrow \pm \langle \hat{\Gamma}_{{\bf k};\mathrm{I}/\mathrm{II}}^{\dag} ( \partial_{\alpha} \hat{\sigma}_{3}\hat{E}_{{\bf k}})^{(\mathrm{I}/\mathrm{II})}  \hat{\Gamma}_{{\bf k};\mathrm{I}/\mathrm{II}} \rangle
\nonumber
\\
&
= \pm \langle \hat{\Gamma}_{{\bf k};\mathrm{I}/\mathrm{II}}^{\dag} ( \partial_{\alpha}\mathrm{diag}[\epsilon^{(\mathrm{I}/\mathrm{II})}_{{\bf k};+}, -\epsilon^{(\mathrm{I}/\mathrm{II})}_{{\bf k};-}] ) \hat{\Gamma}_{{\bf k};\mathrm{I}/\mathrm{II}} \rangle,
\nonumber
\end{align}
where by the $\rightarrow$ we picked only the diagonal part of $\hat{T}_{{\bf k}}^{\dag}(\partial_{\alpha}\hat{H}_{\bf k})\hat{T}_{{\bf k}} = \partial_{\alpha} \hat{\sigma}_{3}\hat{E}_{\bf k} + \hat{{\cal A}}_{\alpha;{\bf k}}\hat{E}_{\bf k}- \hat{E}_{\bf k}\hat{{\cal A}}_{\alpha;{\bf k}}$, where $\hat{{\cal A}}_{\alpha;{\bf k}} = \hat{T}_{{\bf k}}^{\dag} \hat{\sigma}_{3}\partial_{\alpha}\hat{T}_{{\bf k}}$, and the diagonal part is the first term after the equality sign (for example, see Ref. \cite{ZyuzinKovalevPRL2016}). 
For the $\mathrm{I}$ block we have
\begin{align}
\left( j_{{\bf k};\alpha}^{\mathrm{s}}\right)^{(\mathrm{I})}
& 
= ( \partial_{\alpha}\epsilon^{(\mathrm{I})}_{{\bf k};+} )
\langle  c^{\dag}_{{\bf k};1} c_{{\bf k};1}  \rangle 
-
( \partial_{\alpha}\epsilon^{(\mathrm{I})}_{{\bf k};-} )
\langle c_{-{\bf k};2}c_{-{\bf k};2}^{\dag} \rangle
\nonumber
\\
&
=
 ( \partial_{\alpha}\epsilon^{(\mathrm{I})}_{{\bf k};+} )g(\epsilon^{(\mathrm{I})}_{{\bf k};+})
 -
 ( \partial_{\alpha}\epsilon^{(\mathrm{I})}_{{\bf k};-} ) \left[  1+g(\epsilon^{(\mathrm{I})}_{{\bf k};+}) \right]
 \nonumber
 \\
 &
 = ( \partial_{\alpha}\epsilon^{(\mathrm{I})}_{{\bf k};+} )\mathrm{coth}\left( \frac{\epsilon^{(\mathrm{I})}_{{\bf k};+}}{2T} \right).
\end{align}
While for the $\mathrm{II}$ block it is
\begin{align}
\left( j_{{\bf k};\alpha}^{\mathrm{s}}\right)^{(\mathrm{II})}
& 
= - ( \partial_{\alpha}\epsilon^{(\mathrm{II})}_{{\bf k};+} )
\langle  c^{\dag}_{{\bf k};2} c_{{\bf k};2}  \rangle 
+
( \partial_{\alpha}\epsilon^{(\mathrm{II})}_{{\bf k};-} )
\langle c_{-{\bf k};1}c_{-{\bf k};1}^{\dag} \rangle
\nonumber
\\
&
=
-
 ( \partial_{\alpha}\epsilon^{(\mathrm{II})}_{{\bf k};+} )g(\epsilon^{(\mathrm{II})}_{{\bf k};+})
 +
 ( \partial_{\alpha}\epsilon^{(\mathrm{II})}_{{\bf k};-} ) \left[  1+g(\epsilon^{(\mathrm{II})}_{{\bf k};+}) \right]
 \nonumber
 \\
 &
 =- ( \partial_{\alpha}\epsilon^{(\mathrm{II})}_{{\bf k};+} )\mathrm{coth}\left( \frac{\epsilon^{(\mathrm{II})}_{{\bf k};+}}{2T} \right).
\end{align}
Therefore, 
$
 j_{{\bf k};\alpha}^{\mathrm{s}}= ( \partial_{\alpha}\epsilon^{(\mathrm{I})}_{{\bf k};+} )\mathrm{coth}\left( \frac{\epsilon^{(\mathrm{I})}_{{\bf k};+}}{2T} \right) - (\mathrm{I} \rightarrow \mathrm{II}).
$
Thus, upon integrating the expression above over the Brillouin zone we obtain the Eq. (10) of the Main Text.
Presence of factor of $1$ in $\coth(z/2T) = 1 + 2g(z)$ in the expression for the spin current is due to the quantum contribution. 
This factor of $1$ occurred in Eqs. (\ref{correlator1}) and (\ref{correlator2}), and is simply stating that at $T =0$ expectation value $\langle c_{-{\bf k};1/2}c_{-{\bf k};1/2}^{\dag}\rangle = 1$. We don't see any physical or mathematical reasons to neglect these averages when calculating spin current (see a comment Ref. (48) in the Main Text). Note that the spin operator doesn't have any quantum contributions,
\begin{align}
S_{i} &= -S^{z}_{i;\mathrm{R}} - S^{z}_{i;\mathrm{b}} = a_{i}^{\dag}a_{i} - b_{i}^{\dag}b_{i} 
\nonumber
\\
&= \frac{1}{2}\left( a_{i}^{\dag}a_{i}  - b_{i}^{\dag}b_{i} +  a_{i}a^{\dag}_{i} - b_{i} b^{\dag}_{i} \right),
\end{align}
where $i$ is the unit cell number, and there are no peculiarities in the spin density at $T=0$.
In case of ferromagnetic order we would have had $g(z)$ instead of $\coth(z/2T)$ in the expression for the spin current. 
Also recall, that the expression of the supercurrent in superconductors comes with $\tanh(z/2T)$.

\section{Derivation of the equilibrium spin current within Keldysh field integral formalism}

In Keldysh formalism, the Hamiltonian and the fields are split into forward (+) and backward (-) contours, 
\begin{align}
\int_{C} \bar{\Psi}\hat{H}\Psi = \int_{-\infty}^{+\infty} \bar{\Psi}_{+}\hat{H}_{+}\Psi_{+} -  \int_{-\infty}^{+\infty} \bar{\Psi}_{-}\hat{H}_{-}\Psi_{-}
\end{align}
Let us first study only the $\mathrm{I}$ sector (for simplicity we omit but keep in mind an index $\mathrm{I}$ on each field and in the Hamiltonian) described by $\hat{\Psi}^{\dag}_{{\bf k};\mathrm{I}} = (a_{\bf k}^{\dag} ,~b_{-\bf k} )$ spinor,
\begin{align}
\hat{H}_{\pm} = SJ \left[
\begin{array}{cc}
3 & -\gamma_{{\bf k},\pm} \\
 -\gamma_{{\bf k},\pm} & 3
\end{array}
\right],
\end{align}
where $\gamma_{{\bf k},\pm} \equiv \gamma_{{\bf k} + {\bf A}_{\pm}}$ contains a vector potential ${\bf A}_{\pm}$ specified on each Keldysh contour. We stress that there is no true vector potential for magnons as magnons have no charge. However, in order to properly define the current, we introduce such a vector potential for the time being. At the end of the calculation, this vector potential will be set to zero, and only effective vector potential due to the Dzyaloshinskii-Moriya interaction will be left.   
Alternatively, we can trace the nature of the Dzyaloshinskii-Moriya vector potential by introducing a vector potential ${\bf A}_{\pm}$ as $\gamma_{{\bf k},\pm} \equiv \gamma_{{\bf k} + {\bf A}_{\mathrm{DMI}}+ {\bf A}_{\pm}}$, where ${\bf A}_{\mathrm{DMI}}$ is the Dzyaloshinskii-Moriya vector potential. It is clear that the classical component of the vector potential, i.e. $ {\bf A}_{\mathrm{cl}} =\left( {\bf A}_{+} + {\bf A}_{-} \right)/2$ can be associated with the ${\bf A}_{\mathrm{DMI}}$. 
Indeed, 
\begin{align}
 {\bf A}_{\mathrm{DMI}}+ {\bf A}_{\pm} = {\bf A}_{\mathrm{DMI}} + {\bf A}_{\mathrm{cl}} \pm {\bf A}_{\mathrm{q}} =   \tilde{{\bf A}}_{\mathrm{DMI}} \pm {\bf A}_{\mathrm{q}} ,
\end{align}
where ${\bf A}_{{\mathrm q}} =\left( {\bf A}_{+} - {\bf A}_{-} \right)/2$ is the quantum component of the vector potential, and  $\tilde{{\bf A}}_{\mathrm{DMI}} =  {\bf A}_{\mathrm{DMI}} + {\bf A}_{\mathrm{cl}} $. 
It is the quantum component is all we need to properly define the magnon current.
The current of the $\mathrm{I}$ sector of magnons is
\begin{align}
 \langle {\bf j}\rangle = \frac{i}{2}\left. \frac{\partial Z}{\partial {\bf A}_{\mathrm q}} \right|_{{\bf A}_{\mathrm q} = 0},
\end{align}
where $Z$ is the partition function.
Since we are interested in effects that are linear in quantum component of the vector potential, it is enough to approximate
\begin{align}
\gamma_{{\bf k},\pm} \approx \gamma_{{\bf k}} \pm \left({\bm \partial} \gamma_{{\bf k} }\right) \cdot {\bf A}_{\mathrm{q}},
\end{align}
where ${\bf A}_{\mathrm{DMI}}$ is kept in the definition of $\gamma_{{\bf k}}$.
Then
\begin{align}
\hat{H}_{\pm} \approx \hat{H}  - SJ 
\left[\begin{array}{cc} 0 & {\bm \partial} \gamma_{\bf k} \\ {\bm \partial} \gamma_{\bf k}^{*} & 0  \end{array}\right] \cdot \left(\pm {\bf A}_{\mathrm{q}}\right),
\end{align}
The boson fields are defined as 
\begin{align}
\Psi_{\mathrm{cl}/\mathrm{q}} = \frac{1}{\sqrt{2}} \left( \Psi_{+} \pm \Psi_{-} \right), ~~~~ 
\bar{\Psi}_{\mathrm{cl}/\mathrm{q}} = \frac{1}{\sqrt{2}} \left( \bar{\Psi}_{+} \pm \bar{\Psi}_{-} \right).
\end{align}
Then, as usual, 
\begin{align}
\bar{\Psi}_{+}\Psi_{+} - \bar{\Psi}_{-}\Psi_{-} = \bar{\Psi}_{\mathrm{cl}}\Psi_{\mathrm{q}} - \bar{\Psi}_{\mathrm{q}}\Psi_{\mathrm{cl}} ,
\end{align}
\begin{align}
\bar{\Psi}_{+}\Psi_{+} {\bf A}_{\mathrm{q}}  + \bar{\Psi}_{-}\Psi_{-}{\bf A}_{\mathrm{q}} 
=  
\left( \bar{\Psi}_{\mathrm{cl}}\Psi_{\mathrm{cl}} + \bar{\Psi}_{\mathrm{q}}\Psi_{\mathrm{q}} \right) {\bf A}_{\mathrm{q}} .
\end{align}
The unperturbed part of the action is then
\begin{align}
iS_{0} = 
i \int_{-\infty}^{+\infty} dt \int_{\bf k} \hat{\bar{\Psi}} \left[\begin{array}{cc} 0 & \left( \hat{G}^{-1} \right)^{\mathrm{A}} \\ \left( \hat{G}^{-1} \right)^{\mathrm{R}} & \left( \hat{G}^{-1} \right)^{\mathrm{K}}  \end{array}  \right]_{\mathrm{K}}\hat{\Psi}
\equiv 
 i \int_{-\infty}^{+\infty} dt \int_{\bf k} \hat{\bar{\Psi}} \hat{G}^{-1} \hat{\Psi},
\end{align}
where subindex $\mathrm{K}$ on the matrix denotes the Keldysh space, and $\mathrm{R/A}$ is the retarded/advanced parts of the Green function, and $\mathrm{K}$ is the Keldysh part.
The hat $\hat{}$ corresponds to the space of sublattices (SL).
The perturbation written in Keldysh (K) and sublattice (SL) spaces is 
\begin{align}
iS_{A} = i \int_{-\infty}^{+\infty} dt \int_{\bf k} \hat{\bar{\Psi}} \left[\begin{array}{cc} \hat{{\bf A}}_{\mathrm{q}} & 0 \\ 0 & \hat{{\bf A}}_{\mathrm{q}}  \end{array}  \right]_{\mathrm{K}} \otimes  
\left[\begin{array}{cc} 0 & -SJ{\bm \partial} \gamma_{\bf k} \\ -SJ {\bm \partial} \gamma_{\bf k}^{*} & 0  \end{array}\right]_{\mathrm{SL}} \hat{\Psi} 
\equiv 
 i \int_{-\infty}^{+\infty} dt \int_{\bf k} \hat{\bar{\Psi}} \hat{V}\hat{\Psi} ,
\end{align}
then 
\begin{align}
iS = iS_{0} + iS_{A} =  i \int_{-\infty}^{+\infty} dt \int_{\bf k}  \hat{\bar{\Psi}}\left[\hat{G}^{-1} + \hat{V} \right] \hat{\Psi} ,
\end{align}
upon integrating over the boson fields, 
\begin{align}
Z = e^{-\mathrm{Tr}\ln\left[ 1+ \hat{G} \hat{V}\right]},
\end{align}
where 
\begin{align}
\hat{G} = \left[\begin{array}{cc} \hat{G}^{\mathrm{K}} & \hat{G}^{\mathrm{R}} \\ \hat{G}^{\mathrm{A}} & 0  \end{array}  \right]_{\mathrm{K}},
\end{align}
and the Keldysh part of the Green function is defined from the kinetic equation, $\hat{G}^{\mathrm{K}}\left( \hat{G}^{-1} \right)^{\mathrm{A}} + \left( \hat{G}^{-1} \right)^{\mathrm{R}}\hat{G}^{\mathrm{K}} = 0 $, and obtained to be
\begin{align}
\hat{G}^{\mathrm{K}}(\epsilon) = \hat{G}^{\mathrm{R}}(\epsilon) {\cal F}_{\epsilon} - {\cal F}_{\epsilon}\hat{G}^{\mathrm{A}}(\epsilon),
\end{align}
where ${\cal F}_{\epsilon} = \coth \left( \frac{\epsilon}{2T} \right)$ is the boson distribution function, and $T$ is the temperature. Retarded/advanced Green function is 
\begin{align}\label{SMGreenAFM}
G^{\mathrm{R}/\mathrm{A}}(\epsilon;{\bf k})
&
 =\left[
\begin{array}{cc}
 \epsilon \pm i0 - 3SJ & SJ\gamma_{{\bf k}} \\
 SJ\gamma_{{\bf k}} &  -\epsilon \pm i0 - 3SJ
\end{array}
\right]_{\mathrm{SL}}^{-1}
\\
&
=-\frac{1}{(\epsilon \pm i0)^2 - (SJ)^2[9 - \vert\gamma_{{\bf k}} \vert^2]}
\left[
\begin{array}{cc}
-(\epsilon \pm i0) - 3SJ & -SJ\gamma_{{\bf k}} \\
-SJ\gamma_{{\bf k}}^{*} & (\epsilon \pm i0) - 3SJ 
\end{array}
\right]_{\mathrm{SL}}
\end{align}
The current is
\begin{align}
\langle {\bf j}\rangle
&
 = -SJ \frac{i}{2} \mathrm{Tr} \hat{G}^{\mathrm{K}}(\epsilon;{\bf k})  
\left[\begin{array}{cc} 0 & {\bm \partial} \gamma_{\bf k} \\ {\bm \partial} \gamma_{\bf k}^{*} & 0  \end{array}\right]_{\mathrm{SL}}
\\
&
=
- (SJ)^2  \frac{i}{2} \int_{\bf k}\int_{\epsilon}  \left[ \gamma_{\bf k}^{*} {\bm \partial} \gamma_{\bf k} + \gamma_{\bf k} {\bm \partial} \gamma_{\bf k}^{*}\right]
\left[ \frac{{\cal F}_{\epsilon}}{(\epsilon + i0)^2 - (SJ)^2[9 - \vert\gamma_{{\bf k}} \vert^2]} - 
\frac{{\cal F}_{\epsilon}}{(\epsilon - i0)^2 - (SJ)^2[9 - \vert\gamma_{{\bf k}} \vert^2]}
\right]
\\
&
= -
\frac{SJ}{2}  \frac{i}{2} \int_{\bf k}\int_{\epsilon}  \left[ \gamma_{\bf k}^{*} {\bm \partial} \gamma_{\bf k} + \gamma_{\bf k} {\bm \partial} \gamma_{\bf k}^{*}\right]
\frac{{\cal F}_{\epsilon}}{\sqrt{9 - \vert\gamma_{{\bf k}} \vert^2}}
\left[  \frac{1}{(\epsilon + i0) - SJ\sqrt{9 - \vert\gamma_{{\bf k}} \vert^2}} - \frac{1}{(\epsilon + i0) + SJ\sqrt{[9 - \vert\gamma_{{\bf k}} \vert^2]}}
- 
\mathrm{c.c.}
\right]
\\
&
= -
SJ \frac{\pi}{2}  \int_{\bf k}\int_{\epsilon}  \left[ \gamma_{\bf k}^{*} {\bm \partial} \gamma_{\bf k} + \gamma_{\bf k} {\bm \partial} \gamma_{\bf k}^{*}\right]
\frac{{\cal F}_{\epsilon}}{\sqrt{9 - \vert\gamma_{{\bf k}} \vert^2}} \left[ \delta(\epsilon - SJ\sqrt{9 - \vert\gamma_{{\bf k}} \vert^2}) - \delta(\epsilon + SJ\sqrt{9 - \vert\gamma_{{\bf k}} \vert^2}) \right]
\\
&
= -
SJ\frac{1}{2}\int_{\bf k}\frac{{\bm \partial}\vert \gamma_{\bf k} \vert^2}{\sqrt{9 - \vert\gamma_{{\bf k}} \vert^2}} \coth\left( \frac{SJ\sqrt{9 - \vert\gamma_{{\bf k}} \vert^2}}{2T}\right)
\\
&
=
SJ\int_{\bf k}{\bm \partial}\sqrt{9 - \vert\gamma_{{\bf k}} \vert^2} \coth\left( \frac{SJ\sqrt{9 - \vert\gamma_{{\bf k}} \vert^2}}{2T}\right),
\end{align}
where $\int_{\epsilon} \equiv \int \frac{d\epsilon}{2\pi}$ and $\int_{{\bf k}} \equiv \int_{\mathrm{BZ}} \frac{d^{n}k}{(2\pi)^{n}}$ is the integral over the Brillouin zone and $n$ is the dimension of the system (it is $n=2$ in our case).
This is the spin current contribution of only the $\mathrm{I}$ block, i.e. $\langle {\bf j}\rangle$ is $\langle {\bf j}\rangle \equiv \langle {\bf j}^{\mathrm{I}}\rangle$. If we do the same steps for the $\mathrm{II}$ block, 
we will recover the expression for the spin current outlined in Eq. (10) of the Main Text. Indeed, the vector potential enters the $\mathrm{II}$ block with the opposite sign as compared to the $\mathrm{I}$ block.
Then,
\begin{align}
\langle {\bf j}^{\mathrm{s}}\rangle = \langle {\bf j}^{\mathrm{I}}\rangle - \langle {\bf j}^{\mathrm{II}}\rangle.
\end{align}

We would have obtained the same expression for the equilibrium spin current if we used the Matsubara formalism. This is understood by recalling that there is an identity
\begin{align}
\pi i T \sum_{n} g(\epsilon_{n}) = \frac{1}{4} \int d\epsilon \coth\left( \frac{\epsilon}{2T}\right) \left[ g^{\mathrm{R}}(\epsilon) - g^{\mathrm{A}}(\epsilon) \right],
\end{align}
where $g^{\mathrm{R}/\mathrm{A}}(\epsilon) = g(\epsilon \pm i0)$, and $\epsilon_{n} = 2\pi T n$ is the boson Matsubara frequency.

We note that there was never an instance in the presented calculation at which we had to neglect by hand what only appear to be the negative-frequency magnon branches.
For example, these are $\epsilon_{{\bf k};-}^{(\mathrm{II})}$ and $\epsilon_{{\bf k};-}^{(\mathrm{I})}$ in Eq. (3) of the Main Text.
There is a common belief that one has to omit these branches by hand. 
As we showed here, all branches of magnons contribute to the physical observables.

\section{Ground state current}
Having understood in details the diagonal basis and introduction of the vector potential in to the Hamiltonian, it is now instructive to discuss the nature of the quantum component of the magnon equilibrium spin current.

Let us again study only the $\mathrm{I}$ sector described by $\hat{\Psi}^{\dag}_{{\bf k};\mathrm{I}} = (a_{\bf k}^{\dag} ,~b_{-\bf k} )$ spinor,
\begin{align}
\hat{H}_{\mathrm{I};{\bf k}} = SJ \left[
\begin{array}{cc}
3 & -\gamma_{{\bf k}} \\
 -\gamma_{{\bf k}} & 3
\end{array}
\right],
\end{align}
where $\gamma_{{\bf k}} \equiv \gamma_{{\bf k} + {\bf A}_{\mathrm{DMI}} + {\bf A} }$ contains a vector potential ${\bf A}$ apart from the ${\bf A}_{\mathrm{DMI}}$. We needed ${\bf A}$ only to define the current (in other words, ${\bf A}$ is the generator of the current), at the end of the calculation ${\bf A}$ will be set to zero, while ${\bf A}_{\mathrm{DMI}}$ will be kept as it is.
We define the current as 
\begin{align}
{\bf j}^{\mathrm{I}} = \left. \hat{\Psi}^{\dag}_{{\bf k};\mathrm{I}} \frac{\partial }{\partial {\bf A}} \hat{H}_{\mathrm{I};{\bf k}} \hat{\Psi}_{{\bf k};\mathrm{I}} \right|_{{\bf A}= 0}
\rightarrow
\left. \frac{\partial }{\partial {\bf A}}\hat{\Psi}^{\dag}_{{\bf k};\mathrm{I}} \hat{H}_{\mathrm{I};{\bf k}} \hat{\Psi}_{{\bf k};\mathrm{I}} \right|_{{\bf A}= 0},
\end{align}
where by the $\rightarrow$ we picked only the diagonal elements of the current. Off-diagonal elements are known to result in magnon spin Nernst or magnon thermal Hall effects, but they don't contribute to the magnon equilibrium spin current.
The eigenvalue equation is $\hat{H}_{\bf k}\psi_{\bf k} = \hat{\sigma}_{3}  \hat{E}_{\bf k}\psi_{\bf k} $.

Then, in the diagonal basis we get for the current
\begin{align}
\int_{\bf k}\left.\frac{\partial }{\partial {\bf A}}\hat{\Psi}^{\dag}_{{\bf k};\mathrm{I}} \hat{H}_{\mathrm{I};{\bf k}} \hat{\Psi}_{{\bf k};\mathrm{I}} \right|_{{\bf A}= 0}
&
= 
\int_{\bf k}
\left. \hat{\Gamma}_{{\bf k};\mathrm{I}}^{\dag}\frac{\partial }{\partial {\bf A}}
\left[\begin{array}{cc} \epsilon^{(\mathrm{I})}_{{\bf k};+} & 0 \\ 0 & -\epsilon^{(\mathrm{I})}_{{\bf k};-} \end{array} \right]\hat{\Gamma}_{{\bf k};\mathrm{I}}\right|_{{\bf A}= 0}
\\
&
=
\left.
\int_{\bf k}\left[ c^{\dag}_{{\bf k};1}c_{{\bf k};1}\frac{\partial }{\partial {\bf A}}\epsilon^{(\mathrm{I})}_{{\bf k};+}
+c_{-{\bf k};2}c^{\dag}_{-{\bf k};2}\frac{\partial }{\partial {\bf A}}\epsilon^{(\mathrm{I})}_{{\bf k};+} \right]
\right|_{{\bf A}= 0}
\\
&
=
\left.
\int_{\bf k}\left[ c^{\dag}_{{\bf k};1}c_{{\bf k};1}\frac{\partial }{\partial {\bf A}}\epsilon^{(\mathrm{I})}_{{\bf k};+}
+ c_{{\bf k};2}^{\dag}c_{{\bf k};2}\frac{\partial }{\partial {\bf A}}\epsilon^{(\mathrm{II})}_{{\bf k};+}
+\frac{\partial }{\partial {\bf A}}\epsilon^{(\mathrm{I})}_{{\bf k};+} \right]
\right|_{{\bf A}= 0}
\end{align}
where we have used  $\epsilon^{(\mathrm{I})}_{{\bf k};-} = -  \epsilon^{(\mathrm{I})}_{{\bf k};+}$ identity and $\epsilon^{(\mathrm{I})}_{-{\bf k};+} = \epsilon^{(\mathrm{II})}_{{\bf k};+}$ true for our particular model.
Similarly we get for the $\mathrm{II}$ block
\begin{align}
\int_{\bf k}\left.\frac{\partial }{\partial {\bf A}}\hat{\Psi}^{\dag}_{{\bf k};\mathrm{II}} \hat{H}_{\mathrm{II};{\bf k}} \hat{\Psi}_{{\bf k};\mathrm{II}} \right|_{{\bf A}= 0}
&
= 
\int_{\bf k}
\left. \hat{\Gamma}_{{\bf k};\mathrm{II}}^{\dag}\frac{\partial }{\partial {\bf A}}
\left[\begin{array}{cc} \epsilon^{(\mathrm{II})}_{{\bf k};+} & 0 \\ 0 & -\epsilon^{(\mathrm{II})}_{{\bf k};-} \end{array} \right]\hat{\Gamma}_{{\bf k};\mathrm{II}}\right|_{{\bf A}= 0}
\\
&
=
\left.
\int_{\bf k}\left[ c^{\dag}_{{\bf k};2}c_{{\bf k};2}\frac{\partial }{\partial {\bf A}}\epsilon^{(\mathrm{II})}_{{\bf k};+}
+c_{-{\bf k};1}c^{\dag}_{-{\bf k};1}\frac{\partial }{\partial {\bf A}}\epsilon^{(\mathrm{II})}_{{\bf k};+} \right]
\right|_{{\bf A}= 0}
\\
&
=
\left.
\int_{\bf k}\left[ c^{\dag}_{{\bf k};2}c_{{\bf k};2}\frac{\partial }{\partial {\bf A}}\epsilon^{(\mathrm{II})}_{{\bf k};+}
+ c_{{\bf k};1}^{\dag}c_{{\bf k};1}\frac{\partial }{\partial {\bf A}}\epsilon^{(\mathrm{I})}_{{\bf k};+}
+\frac{\partial }{\partial {\bf A}}\epsilon^{(\mathrm{II})}_{{\bf k};+} \right]
\right|_{{\bf A}= 0}.
\end{align}
We use $\partial_{A_{\alpha}} \epsilon^{(\mathrm{I})}_{{\bf k};+}= \partial_{k_{\alpha}}\epsilon^{(\mathrm{I})}_{{\bf k};+}$ and $\partial_{A_{\alpha}} \epsilon^{(\mathrm{II})}_{{\bf k};+}= -\partial_{k_{\alpha}}\epsilon^{(\mathrm{II})}_{{\bf k};+}$, where the latter is because $\mathrm{II}$ contains $\gamma_{-{\bf k} + {\bf A}_{\mathrm{DMI}} + {\bf A}}$ structure.
Therefore, 
\begin{align}
\langle\int_{\bf k}\left.\frac{\partial }{\partial A_{\alpha}} \hat{\Psi}^{\dag}_{{\bf k};\mathrm{I}} \hat{H}_{\mathrm{I};{\bf k}} \hat{\Psi}_{{\bf k};\mathrm{I}} \right|_{{\bf A}= 0} \rangle
+
\langle\int_{\bf k}\left.\frac{\partial }{\partial  A_{\alpha}}\hat{\Psi}^{\dag}_{{\bf k};\mathrm{II}} \hat{H}_{\mathrm{II};{\bf k}} \hat{\Psi}_{{\bf k};\mathrm{II}} \right|_{{\bf A}= 0} \rangle = 
\left[1+2g(\epsilon^{(\mathrm{I})}_{{\bf k};+})\right] \partial_{k_{\alpha}}\epsilon^{(\mathrm{I})}_{{\bf k};+} - 
\left[1+2g(\epsilon^{(\mathrm{II})}_{{\bf k};+}) \right] \partial_{k_{\alpha}}\epsilon^{(\mathrm{II})}_{{\bf k};+},
\end{align}
and we recover the magnon equilibrium spin current obtained in Eq. (10) of the Main Text.
The term with $1$ in the square brackets is exactly the ground state current. It appeared from the quantum correction to the ground state energy.

\end{widetext}
\end{document}